\documentclass[reprint,aps,pra,showpacs,floatfix]{revtex4-1}

\usepackage{graphicx,amsmath,amssymb,psfrag,empheq,hyperref,wasysym}
\usepackage{dcolumn}
\usepackage{bm}
\usepackage{epsfig}
\begin{document}

\title{Dilaton Solutions for Laboratory Constraints and Lunar Laser Ranging}

\author{Philippe Brax}
\email{philippe.brax@ipht.fr}
\affiliation{Institut de Physique Th\'{e}orique, Universit\'e Paris-Saclay, CEA, CNRS, F-91191 Gif/Yvette Cedex, France}

\author{Hauke Fischer}
\email{hauke.fischer@tuwien.ac.at}
\affiliation{Atominstitut, Technische Universit\"at Wien, Stadionallee 2, A-1020 Wien, Austria}

\author{Christian K\"ading}
\email{christian.kaeding@tuwien.ac.at}
\affiliation{Atominstitut, Technische Universit\"at Wien, Stadionallee 2, A-1020 Wien, Austria}

\author{Mario Pitschmann}
\email{mario.pitschmann@tuwien.ac.at}
\affiliation{Atominstitut, Technische Universit\"at Wien, Stadionallee 2, A-1020 Wien, Austria}

\begin{abstract}
We derive approximate analytical solutions to the environment-dependent dilaton field theory equations in the presence of a one or two mirror system or a sphere. The one-dimensional equations of motion are integrated for each system. The solutions obtained herein can be applied to \textit{q}BOUNCE experiments, neutron interferometry and for the calculation of the dilaton field induced "Casimir force" in the \textsc{Cannex} experiment as well as for Lunar Laser Ranging. They are typical of the Damour-Polyakov screening mechanism whereby deviations from General Relativity are suppressed by a vanishingly small direct coupling of the dilaton to matter in dense environments. 
\end{abstract}


\maketitle


\section{Introduction}

The accelerated expansion of the Universe may require the introduction of additional degrees of freedom  (see \cite{Joyce:2014kja} for a recent review). Such new degrees of freedom, in particular light scalars, are theoretically well motivated irrespective of their role for the acceleration of the expansion of the Universe. If they exist in Nature, they must either only be feebly coupling to other matter or appear in some screened form in order to
prevent detection in all past experiments and observations involving scalar fifth forces. A number of screening mechanisms exist \cite{Joyce:2014kja}, the chameleon \cite{Khoury:2003rn,Khoury:2003aq,Brax:2004qh} and Damour-Polyakov mechanisms \cite{Damour:1994zq}, the K-mouflage \cite{Babichev:2009ee,Brax:2012jr,Brax:2014wla} and Vainshtein ones \cite{Vainshtein:1972sx}, allowing such hypothetical  fields to remain unseen in local tests of gravity. 

In the companion paper \cite{Pitschmann:2018aa}, gravity resonance spectroscopy \cite{Abele:2009dw, Jenke:2011zz}, a Casimir force experiment \cite{Sedmik:2018kqt} and Lunar Laser Ranging (LLR) \cite{nordtvedt2001lunar} are used for the first time to put new bounds on the environment-dependent dilaton \cite{Brax:2010gi,Brax:2011ja}. The LLR bounds have already been discussed for chameleon \cite{PhysRevD.97.104044} and symmetron scalar fields \cite{PhysRevD.84.103521}. The dilaton model conjugates two ingredients. First of all, it involves an exponentially decreasing potential as expected in the strong coupling limit of string theory \cite{Gasperini:2001pc,Damour:2002nv,Damour:2002mi} and second it is the simplest realisation of the least coupling principle as advocated in \cite{Damour:1994zq}. Hence this model is a prime example, with the symmetron, of a model subject to the Damour-Polyakov screening mechanism \cite{PhysRevD.82.063519}. 
The experimental analysis in \cite{Pitschmann:2018aa} depends heavily on the field profile of dilatons in both the one mirror or two mirror setups, where the field is either present over an infinite plane of high density or is confined between two such parallel planes. This is the purpose of this paper to provide a detailed analysis of the environment-dependent dilaton in realistic situations which can be tested using current experiments. For instance the analysis of the  LLR results  as well as the screening of neutrons in Q-bounce experiments employ the dilaton solution around and inside  a sphere as derived here.
For a similar study applied to chameleon field theories see \cite{Ivanov:2016rfs}.  The symmetron field equations have been studied in a similar fashion in \cite{Brax:2017hna, Pitschmann:2020ejb}. Testing screened models can be tackled from the laboratory to cosmological scales as reviewed in \cite{Brax:2021wcv}. 

In section \ref{sec:1} we will provide some background information on dilatons, which will provide the relevant definitions for the field theory analysis. We consider a two-dimensional parameter subspace of the dilaton model, which is of special cosmological significance as noted in section \ref{sec:1b}. Then, in section \ref{sec:2} approximative solutions for the one mirror case will be derived, while in section \ref{sec:3} the corresponding two mirror solutions will be given. In section \ref{sec:4} approximate dilaton solutions are derived for a spherical source. Section \ref{sec:5} provides relevant information on the \textit{q}BOUNCE experiment, where the dilaton-induced resonance frequency shift for the case of a single mirror has also been summarized for a large range of parameters. In section \ref{sec:6} the induced pressure in Casimir experiments due to the dilaton field between two mirrors of the experimental setup is derived, while in section \ref{sec:7} bounds due to LLR are given. A conclusion in section \ref{sec:8} will be followed by Appendix \ref{sec:A}, in which the precession of the lunar perigee induced by fifth forces is obtained.

\section{Background}\label{sec:1}

Following \cite{Brax:2018iyo}, the dilaton potential is given by 
\begin{align}
   V(\phi) = V_0\,e^{-\lambda\phi/m_\text{Pl}}\>,
\end{align}
where $V_0$ is an energy scale related to the dark energy of the Universe and $\lambda$ a numerical constant. This potential corresponds to the string theory dilaton potential in the strong coupling limit \cite{Damour:1994zq,Damour:2002nv}. Together with the coupling to matter this induces an effective potential  
\begin{align}
   V_\text{eff}(\phi; \rho) = V(\phi) + A(\phi)\,\rho\>,
\end{align}
where for the environment-dependent dilaton we have \cite{Brax:2010gi}
\begin{align}\label{cf}
   A(\phi) = 1 + \frac{A_2}{2m_\text{Pl}^2}\,\phi^2\>,
\end{align}
and hence
\begin{align}\label{EffP}
  V_\text{eff}(\phi; \rho) = V_0\,e^{-\lambda\phi/m_\text{Pl}} +  \frac{A_2\rho}{2m_\text{Pl}^2}\,\phi^2\>.
\end{align}
Here, we have neglected an additional term $\rho$, which does not affect the equations of motion. 
The minimum value $\phi_\rho$ in the presence of a density $\rho$ is given by $V_{\text{eff},\phi}(\phi; \rho)\big|_{\phi=\phi_\rho}=0$ and reads
\begin{align}\label{PHIVV}
  \phi_\rho = \frac{m_\text{Pl}}{\lambda}\,W\bigg(\frac{\lambda^2V_0}{A_2\rho}\bigg)
\end{align}
with the Lambert $W$-function
\begin{align}
   W(x) &= \sum_{n=1}^\infty\frac{(-n)^{n-1}}{n!}\,x^n \nonumber\\
   &= x - x^2 + \frac{3}{2}\,x^3 - \frac{8}{3}\,x^4 + \frac{125}{24}\,x^5 - \frac{54}{5}\,x^6 + \ldots\>.
\end{align}
For large arguments the approximative relation holds:
\begin{align}
   W(x) \simeq \ln x\>.
\end{align}
The mass $\mu_\rho$ of the quantum fluctuation is therefore
\begin{align}\label{mass}
   \mu_\rho &= \sqrt{V_{\text{eff},\phi\phi}(\phi_\rho; \rho)} \nonumber\\
   &= \frac{1}{m_\text{Pl}}\sqrt{\lambda^2V_0\,e^{-\lambda\phi_\rho/m_\text{Pl}} + A_2\rho}\>.
\end{align}
We employ the metric signature ($+---$), for which the stress-energy tensor of a scalar field is 
\begin{align}\label{SET}
   T_{\mu\nu}^\phi &= \partial_\mu\phi\,\partial_\nu\phi - g_{\mu\nu}\Big(\frac{1}{2}\,\partial_\alpha\phi\,\partial^\alpha\phi - V(\phi; \rho)\Big)\>,
\end{align}
while the equations of motion read
\begin{align}\label{GEOM}
   \Box\phi + V_{\text{eff},\phi}(\phi; \rho) = 0\>
\end{align}
in the presence of matter.
\section{The "Cosmological" Dilaton}\label{sec:1b}

The parameter space of the dilaton model is 3-dimensional ($V_0$, $\lambda$, $A_2$). Since we are interested only in "cosmological" dilatons, having significance in the 
cosmological domain, we consider only the 2-dimensional parameter subspace, for which 
\begin{align}\label{V0VEFF}
   V_\text{eff}(\phi_V; \rho_V) = 3\Omega_{\Lambda0}m_\text{Pl}^2H_0^2 = 6.73\times10^{-34}\>\text{MeV}^4\>,
\end{align}
where $\phi_V$, $\rho_V$ are the corresponding vacuum values and the density parameter $\Omega_{\Lambda0} \sim 0.73$. This choice neglects the possible instability of the potential under radiative corrections. We take the parameter $V_0$ as a function of $A_2$ and $\lambda$ such that Eq.~(\ref{V0VEFF}) is obeyed.
It is important to notice  that the coupling function involves the strong coupling scale
\begin{equation} 
M= \frac{m_{\rm Pl}}{\sqrt A_2}.
\end{equation}
In the spirit of an effective field theory expansion of the coupling function, one must require that $\phi/M \ll 1$ to guarantee that terms of higher order in $\phi$ can be neglected. Similarly the exponential potential involves the scale
\begin{equation}
\Lambda = \frac{m_{\rm Pl}}{\lambda}
\end{equation}
and one must make sure that typically $\phi/\Lambda \gg 1$ to guarantee that higher order string corrections in $e^{-n \phi/\Lambda}$ can be neglected. We will make sure that these conditions are satisfied in the following.

\begin{figure}[ht!]
\centering
\includegraphics[width=\linewidth]{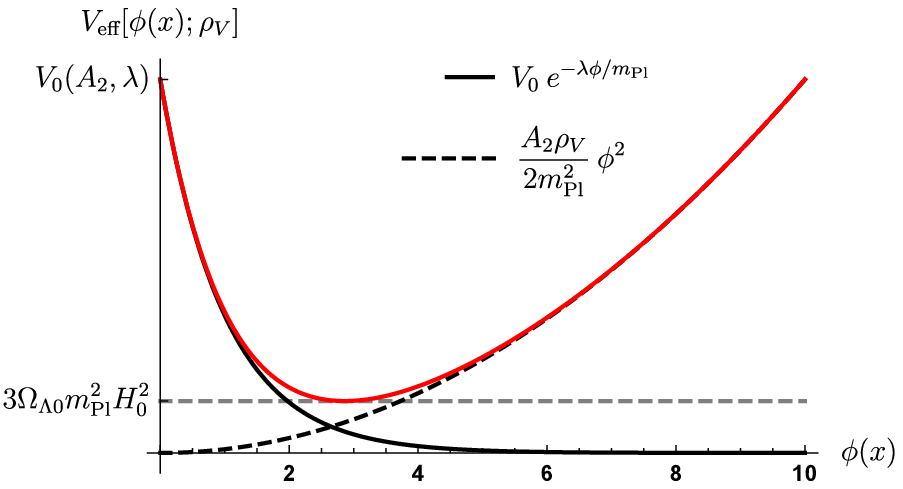} 
\includegraphics[width=\linewidth]{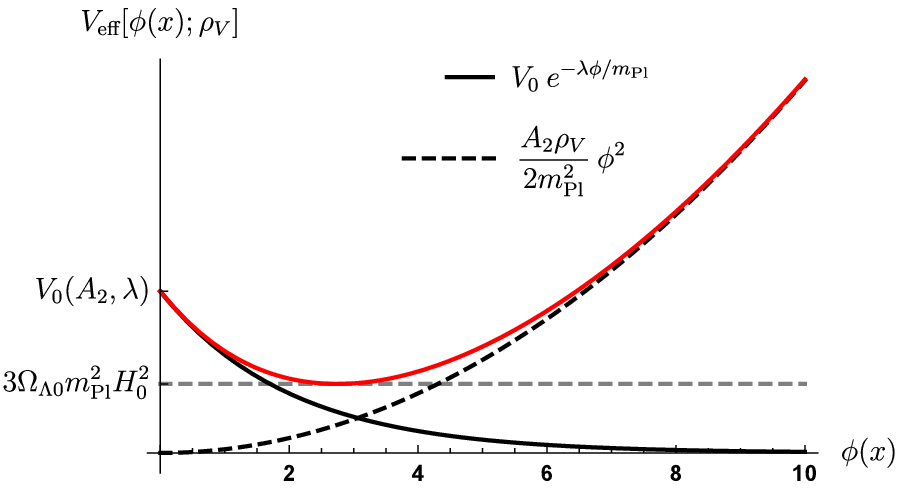} 
\caption{The effective potential in Eq.~(\ref{EffP}) is plotted with its components for two different sets of parameter values for $\lambda$ and $A_2$, while the parameter $V_0$ varies as a function of $A_2$ and $\lambda$ such that the effective potential of the minimum in vacuum equals $3\Omega_{\Lambda0}m_\text{Pl}^2H_0^2$.}
\label{Fig:EffPot}
\end{figure}

It is straightforward to show that the condition (\ref{V0VEFF}) leads to 
\begin{align}\label{V0AL}
   V_0(A_2, \lambda) &= \frac{A_2\rho_V}{\lambda^2}\left(\sqrt{1 + \frac{2\lambda^2}{A_2\rho_V}\,3\Omega_{\Lambda0}m_\text{Pl}^2H_0^2} - 1\right) \nonumber\\
   &\quad\times\exp\left\{\sqrt{1 + \frac{2\lambda^2}{A_2\rho_V}\,3\Omega_{\Lambda0}m_\text{Pl}^2H_0^2} - 1\right\}\>.
\end{align}
The dilaton parameter space becomes effectively 2-dimensional ($\lambda$, $A_2$) simplifying also the representation of experimental constraints.
For illustrative purposes, in Fig.~\ref{Fig:EffPot} the effective potential with its components is depicted for two different sets of parameter values for $\lambda$ and $A_2$. 

Using Eq.~(\ref{V0AL}) in Eq.~(\ref{mass}) the mass $\mu_\rho$ may be expressed as a function of the parameters $\lambda$ and $A_2$ as well
\begin{align}
   \mu_\rho &= \frac{\sqrt{A_2\rho}}{m_\text{Pl}}\,\Bigg\{1 + W\Bigg[\frac{\rho_V}{\rho}\left(\sqrt{1 + \frac{2\lambda^2}{A_2\rho_V}\,3\Omega_{\Lambda0}m_\text{Pl}^2H_0^2} - 1\right) \nonumber\\
   &\quad\times\exp\left(\sqrt{1 + \frac{2\lambda^2}{A_2\rho_V}\,3\Omega_{\Lambda0}m_\text{Pl}^2H_0^2} - 1\right)\Bigg]\Bigg\}^{1/2}\>. 
\end{align}

With these analytical expressions we can perform the limits $A_2\to\infty$, $\lambda\to\infty$, $A_2\to0$ or $\lambda\to0$ and study the behavior along curves $\lambda\propto\sqrt{A_2}$, which all provide some qualitative insight of the relevant parameter region. This behavior is summarized in the following two tables \ref{table:DL} and \ref{table:DL2}.
\begin{table}[!ht]
\centering
\addtolength{\tabcolsep}{2pt}
\renewcommand{\arraystretch}{1.5}
\begin{tabular}{|l|c|c|}
	\hline
	& $\phi_\rho$ &  $\mu_\rho$  \\
	\hline\hline
	$\lim_{A_2\to\infty}$ & $0$ & $\infty$ \\
	\hline	
	$\lim_{\lambda\to\infty}$ & $\displaystyle m_\text{Pl}\,\sqrt{\frac{2}{A_2\rho_V}\,3\Omega_{\Lambda0}m_\text{Pl}^2H_0^2}$ & $\infty$ \\
	\hline	
	$\lim_{A_2\to0}$ & $\infty$ & $0$ \\
	\hline	
	$\lim_{\lambda\to0}$ & $0$ & $\displaystyle\frac{\sqrt{A_2\rho}}{m_\text{Pl}}$ \\
	\hline
	$\lambda\propto\sqrt{A_2}$ & $\sim1/\lambda$ & $\sim\lambda$ \\
	\hline
\end{tabular}
\caption{Here, Eqs.~(\ref{PHIVV}) and (\ref{mass}) are given for different limits.}
\label{table:DL}
\end{table}
\begin{table}[!ht]
\centering
\addtolength{\tabcolsep}{2pt}
\renewcommand{\arraystretch}{1.5}
\begin{tabular}{|l|c|c|}
	\hline
	& $\displaystyle V_0\,e^{-\lambda\phi_\rho/m_\text{Pl}}$ & $\displaystyle\frac{A_2\rho}{2m_\text{Pl}^2}\,\phi_\rho^2$ \\
	\hline\hline
	$\lim_{A_2\to\infty}$ & $3\Omega_{\Lambda0}m_\text{Pl}^2H_0^2$ & $0$ \\
	\hline	
	$\lim_{\lambda\to\infty}$ & $0$ & $\displaystyle\frac{\rho}{\rho_V}\,3\Omega_{\Lambda0}m_\text{Pl}^2H_0^2$ \\
	\hline	
	$\lim_{A_2\to0}$ & $0$ & $\displaystyle\frac{\rho}{\rho_V}\,3\Omega_{\Lambda0}m_\text{Pl}^2H_0^2$ \\
	\hline	
	$\lim_{\lambda\to0}$ & $3\Omega_{\Lambda0}m_\text{Pl}^2H_0^2$ & $0$ \\
	\hline
	$\lambda\propto\sqrt{A_2}$ & const & const \\
	\hline
\end{tabular}
\caption{The limits of the two constituent parts of the effective potential Eq.~(\ref{EffP}) are summarized in this table.}
\label{table:DL2}
\end{table}

We may summarize the findings of the tables as follows. For roughly $\lambda\propto\sqrt{A_2}$ we expect finite limits to be obtainable from experiments  as long as $\lambda$ is not too large since otherwise $\phi_\rho$ decreases, i.e. the dilaton effectively disappears and the interaction range $1/\mu_\rho$ vanishes as well. For $A_2 \gg \lambda$ we expect no limits from experiments since $\phi_\rho$ becomes small.  On the other hand, for $\lambda \gg A_2$ we also do not expect bounds since in this case either $1/\mu_\rho$ decreases without limit or $\phi_\rho$ diverges and cannot act dynamically anymore. Hence, we find that the dilaton has a  physical impact only within a restricted region in the $A_2$, $\lambda$ parameter space. Consequently, all experimental constraints have to lie within this region. As can be seen in table \ref{table:DL}, shorter interaction ranges correspond to larger values of $A_2$ and $\lambda$. Therefore, we expect LLR to probe small parameter values within this region, table top experiments correspondingly larger parameters and collider constraints still larger values. 
Numerical evaluations in the accompanying article \cite{Pitschmann:2018aa} corroborate these findings. Notice that  when $\lambda = \kappa \sqrt A_2$, the validity of the model as an effective field theory is guaranteed when $\kappa \ll 1 $ as the field values must satisfy$\Lambda\ll \phi \ll M$ where $\Lambda = \kappa M \ll M$. These expectations are also confirmed by the actual bounds in \cite{Pitschmann:2018aa}.

\section{one mirror}\label{sec:2}

In this section, we treat the case of a single mirror filling the infinite half-space $z \leq 0$.
The 1-dimensional equation of motion reads (see Eq.~(\ref{GEOM}))
\begin{align}
   \frac{d^2\phi}{dz^2} = V_{\text{eff},\phi}(\phi; \rho)\>.
\end{align}
Multiplication by $\phi'$ and integration with respect to $z$ gives
\begin{align}\label{IEOM}
   \frac{1}{2}\left(\frac{d\phi}{dz}\right)^2 - \frac{1}{2}\left(\frac{d\phi}{dz}\right)^2\bigg|_{z=z_0} = V_\text{eff}(\phi; \rho) - V_\text{eff}(\phi; \rho)\big|_{z=z_0}\>,
\end{align}
with the integration constant $z_0$ and leads to
\begin{align}\label{IIEOM}
   &\int_{\phi_0}^{\phi(z)}\frac{d\phi}{\displaystyle\sqrt{V_\text{eff}(\phi; \rho) - V_\text{eff}(\phi; \rho)\big|_{z=z_0} + \frac{1}{2}\left(\frac{d\phi}{dz}\right)^2\bigg|_{z=z_0}}} \nonumber\\
   &\qquad = \pm\sqrt2\left(z - z_0\right),
\end{align}
where $\phi_0=\phi(z_0)$, the $+$ sign is for $\phi' \geq 0$, while the $-$ sign holds for $\phi' \leq 0$.

\subsection{The Vacuum Region}

First, we consider the case of low density $\rho_V$  corresponding to the medium above the mirror and 
search for a solution that asymptotically for $z \to \infty$ goes as $\phi(z) \to \phi_V$ with
\begin{align}
  \phi_V = \frac{m_\text{Pl}}{\lambda}\,W\bigg(\frac{\lambda^2V_0}{A_2\rho_V}\bigg)\>,
\end{align}
implying $\phi' \to 0$. We find for $z_0=0$ and $z\to\infty$ from Eq.~(\ref{IEOM})
\begin{align}\label{1LD2}
   - \frac{1}{2}\left(\frac{d\phi}{dz}\right)^2\bigg|_{z=0} = V_\text{eff}(\phi_V; \rho_V) - V_\text{eff}(\phi; \rho_V)\big|_{z=0}\>.
\end{align}
Using Eq.~(\ref{1LD2}) in Eq.~(\ref{IIEOM}) gives
\begin{align}
   \int_{\phi_0}^{\phi(z)}\frac{d\phi}{\displaystyle\sqrt{V_\text{eff}(\phi; \rho_V) - V_\text{eff}(\phi_V; \rho_V)}} = \sqrt2\,z\>,
\end{align}
respectively
\begin{align}
   &\int_{\phi_0}^{\phi(z)}\frac{d\phi}{\sqrt{\displaystyle V_0\left(e^{-\lambda\phi/m_\text{Pl}} - e^{-\lambda\phi_V/m_\text{Pl}}\right)+  \frac{A_2\rho_V}{2m_\text{Pl}^2}\left(\phi^2 - \phi_V^2\right)}} \nonumber\\
   &= \sqrt2\,z\>.
\end{align}
Approximating the effective potential around its minimum at the vacuum value $\phi_V$, we find to leading order
\begin{align}
   \frac{1}{\mu_V}\int_{(\phi_V - \phi(z))/m_\text{Pl}}^{(\phi_V - \phi_0)/m_\text{Pl}}\frac{d\tilde\phi}{\tilde\phi} = z
\end{align}
with the mass of the quantum fluctuation in vacuum
\begin{align}\label{muv}
   \mu_V &= \sqrt{V_{\text{eff},\phi\phi}(\phi; \rho_V)}\Big|_{\phi=\phi_V} \nonumber\\
   &= \frac{1}{m_\text{Pl}}\sqrt{\lambda^2V_0\,e^{-\lambda\phi_V/m_\text{Pl}} + A_2\rho_V}\>.
\end{align}
Inverting the relation straightforwardly leads to
\begin{align}
   \phi(z) &= \phi_V + \left(\phi_0 - \phi_V\right)e^{-\mu_Vz}\>.
\end{align}

\subsection{The High Density Region}

Here, we consider the case of high density $\rho_M$ as inside the mirror.
Clearly, for $z\to-\infty$ we have $\phi(z)\to\phi_M$ with
\begin{align}
  \phi_M = \frac{m_\text{Pl}}{\lambda}\,W\bigg(\frac{\lambda^2V_0}{A_2\rho_M}\bigg)\>,
\end{align}
and hence $\phi'\to0$. Therefore, we find for $z_0=0$ and $z\to-\infty$ from Eq.~(\ref{IEOM})
\begin{align}\label{1HD2}
   - \frac{1}{2}\left(\frac{d\phi}{dz}\right)^2\bigg|_{z=0} = V_\text{eff}(\phi_M; \rho_M) - V_\text{eff}(\phi; \rho_M)\big|_{z=0}\>.
\end{align}
Using Eq.~(\ref{1HD2}) in Eq.~(\ref{IIEOM}) gives
\begin{align}
   \int_{\phi_0}^{\phi(z)}\frac{d\phi}{\displaystyle\sqrt{V_\text{eff}(\phi; \rho_M) - V_\text{eff}(\phi_M; \rho_M)}} = \sqrt2\,z\>.
\end{align}
Proceeding analogously to the case of vacuum, we finally obtain
\begin{align}
  \phi(z) &= \phi_M + \left(\phi_0 - \phi_M\right)e^{\mu_Mz}\>,
\end{align}
with the mass of the quantum fluctuation in the mirror 
\begin{align}
   \mu_M = \frac{1}{m_\text{Pl}}\sqrt{\lambda^2V_0\,e^{-\lambda\phi_M/m_\text{Pl}} + A_2\rho_M}\>.
\end{align}

\subsection{Boundary Conditions}

Using the boundary conditions
\begin{align}
   \frac{d\phi}{dz}\bigg|_{z=0_-} = \frac{d\phi}{dz}\bigg|_{z=0_+}\>,
\end{align}
we find
\begin{align}
  \phi_0 = \frac{\mu_V\,\phi_V + \mu_M\,\phi_M}{\mu_V + \mu_M}\>.
\end{align}
The second boundary condition 
\begin{align}
   \phi(0_-) = \phi(0_+)\>,
\end{align}
is trivially satisfied.

\subsection{Final Solution}

\begin{figure}[h]
\centering
\includegraphics[width=\linewidth]{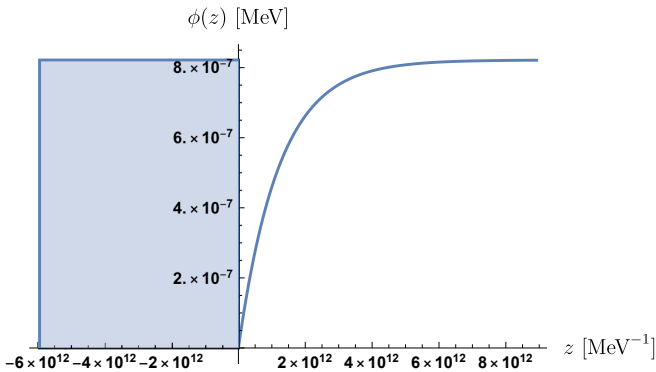} 
\caption{This plot shows the dilaton solution with the blue area depicting the mirror of the experimental setup. We can see that  $\lambda \phi(z) \ll m_\text{Pl} = 1.22\times10^{22}$ MeV is indeed satisfied in this case.}
\label{Fig1}
\end{figure}

Summarising, we obtain the solution
\begin{align}\label{FS1M}
   \phi(z) &= \Theta(+z)\,\big(\phi_V + \left(\phi_0 - \phi_V\right)e^{-\mu_Vz}\big)  \nonumber\\
   &\quad+ \Theta(-z)\,\big(\phi_M + \left(\phi_0 - \phi_M\right)e^{\mu_Mz}\big)\>,
\end{align}
with
\begin{align}
  \mu_V &= \frac{1}{m_\text{Pl}}\sqrt{\lambda^2V_0\,e^{-\lambda\phi_V/m_\text{Pl}} + A_2\rho_V}\>, \nonumber\\
  \mu_M &= \frac{1}{m_\text{Pl}}\sqrt{\lambda^2V_0\,e^{-\lambda\phi_M/m_\text{Pl}} + A_2\rho_M}\>, 
\end{align}
and
\begin{align}\label{eq:phi0}
  \phi_0 = \frac{\mu_V\,\phi_V + \mu_M\,\phi_M}{\mu_V + \mu_M}\>.
\end{align}
A prototype solution with values $A_2 = 10^{35}$, $\lambda = 10^{25}$ and $V_0 = 1.00034\times3\Omega_{\Lambda0}m_\text{Pl}^2H_0^2 = 6.73\times10^{-10}$ meV$^4$, $\rho_V = 10^{-15}$ MeV$^4$ and $\rho_M = 1.082\times10^{-5}$ MeV$^4$ is plotted in Fig.~\ref{Fig1}.

These dilaton solutions will be instrumental in analyzing the results of \textit{q}BOUNCE experiments as outlined in section \ref{sec:5}.

\section{two mirrors}\label{sec:3}

In this section, we treat the case of two parallel infinitely thick mirrors separated at distance $2d$ in $z$-direction with $z=0$ being the center between the two mirrors. 

\subsection{The Vacuum Region}

Here, we consider the case of low density $\rho_V$, as between the mirrors and choose $z_0=0$. Due to the symmetry of the setup, the derivative of the field has to vanish there yielding for Eq.~(\ref{IIEOM})
\begin{align}
   \int_{\phi_0}^{\phi(z)}\frac{d\phi}{\displaystyle\sqrt{V_\text{eff}(\phi; \rho_V) - V_\text{eff}(\phi_0; \rho_V)}} = \sqrt2\,z\>,
\end{align}
where $\phi_0 = \phi(0)$. Approximating the potential around $\phi_0$ yields
\begin{align}
   V_\text{eff}(\phi; \rho_V) - V_\text{eff}(\phi_0; \rho_V) \simeq -\mathfrak{D}_0\left(\phi - \phi_0\right) + \frac{\mu_0^2}{2}\left(\phi - \phi_0\right)^2\>,
\end{align}
with the positive constants
\begin{align}
   \mathfrak{D}_0 &= \frac{\lambda V_0}{m_\text{Pl}}\,e^{-\lambda\phi_0/m_\text{Pl}} - \frac{A_2\rho_V}{m_\text{Pl}^2}\,\phi_0\>,  \nonumber\\
   \mu_0 &= \frac{1}{m_\text{Pl}}\sqrt{\lambda^2V_0\,e^{-\lambda\phi_0/m_\text{Pl}} + A_2\rho_V}\>,
\end{align}
leads to
\begin{align}
   \int_{\phi_0}^{\phi(z)}\frac{d\phi}{\sqrt{\displaystyle -\mathfrak{D}_0\left(\phi - \phi_0\right) + \frac{\mu_0^2}{2}\left(\phi - \phi_0\right)^2}} = -\sqrt2\,|z|\>.
\end{align}
Substituting $\displaystyle x = 1 + \frac{\mu_0^2}{\mathfrak{D}_0}\left(\phi_0 - \phi\right)$ leads to the elementary integral
\begin{align}
   \frac{1}{\mu_0}\int_{1}^{1 + \frac{\mu_0^2}{\mathfrak{D}_0}\,(\phi_0 - \phi(z))}\frac{dx}{\sqrt{\displaystyle x^2 - 1}} = |z|\>.
\end{align}
Integration and some transformations lead to the solution
\begin{align}
  \phi(z) = \phi_0 + \frac{\mathfrak{D}_0}{\mu_0^2}\,\big(1 - \cosh(\mu_0z)\big)\>.
\end{align}

\subsection{The High Density Region}

We can read off the solution inside the mirrors directly from the corresponding solution in the one mirror case Eq.~(\ref{FS1M})
\begin{align}
  \phi(z) &= \phi_M + \left(\phi_d - \phi_M\right)e^{-\mu_M(|z| - d)}\>,
\end{align}
where $\phi_d = \phi(d) = \phi(-d)$ is the field value at the surfaces of the mirrors.

\subsection{Boundary Conditions}

Using the boundary conditions at the mirror surface
\begin{align}
   \phi(d_-) = \phi(d_+)\>,
\end{align}
gives $\phi_d$ as a function of $\phi_0$
\begin{align}
   \phi_d = \phi_0 + \frac{\mathfrak{D}_0}{\mu_0^2}\left(1 - \cosh(\mu_0d)\right)\>,
\end{align}
which together with the second boundary condition 
\begin{align}
   \frac{d\phi}{dz}\bigg|_{z=d_-} = \frac{d\phi}{dz}\bigg|_{z=d_+}
\end{align}
gives an equation, which defines $\phi_0$ implicitly 
\begin{align}
   \phi_M + \frac{\mathfrak{D}_0}{\mu_0\mu_M}\sinh(\mu_0d) = \phi_0 + \frac{\mathfrak{D}_0}{\mu_0^2}\left(1 - \cosh(\mu_0d)\right)
\end{align}
since $\mathfrak{D}_0$ and $\mu_0$ both are functions of $\phi_0$.

\subsection{Final Solution}

\begin{figure}[h]
\centering
\includegraphics[width=\linewidth]{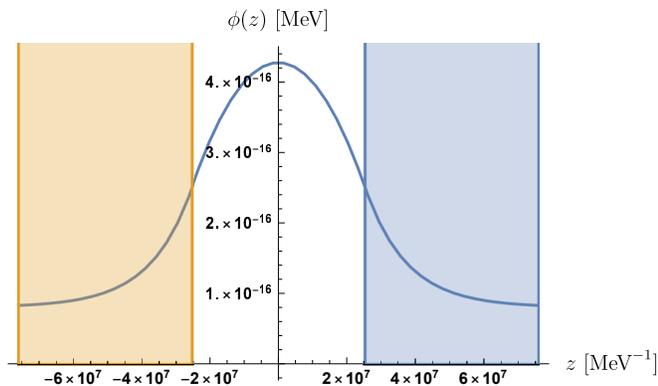} 
\caption{This plot shows the dilaton solution with the blue and orange area depicting the mirrors of the experimental setup. We can see that  $\lambda \phi(z) \ll m_\text{Pl} = 1.22\times10^{22}$ MeV is indeed satisfied in this case.}
\label{Fig2}
\end{figure}

Summarising, we obtain the solution
\begin{align}
   \phi(z) &= \Theta(d - |z|)\left\{\phi_0 + \frac{\mathfrak{D}_0}{\mu_0^2}\,\big(1 - \cosh(\mu_0z)\big)\right\} \nonumber\\
   &\quad + \Theta(|z| - d)\left\{\phi_M + \left(\phi_d - \phi_M\right)e^{-\mu_M(|z| - d)}\right\}\>,
\end{align}
where
\begin{align}
  \mu_M &= \frac{1}{m_\text{Pl}}\sqrt{\lambda^2V_0\,e^{-\lambda\phi_M/m_\text{Pl}} + A_2\rho_M}\>, \nonumber\\
  \mu_0 &= \frac{1}{m_\text{Pl}}\sqrt{\lambda^2V_0\,e^{-\lambda\phi_0/m_\text{Pl}} + A_2\rho_V}\>, \nonumber\\
  \mathfrak{D}_0 &= \frac{\lambda V_0}{m_\text{Pl}}\,e^{-\lambda\phi_0/m_\text{Pl}} - \frac{A_2\rho_V}{m_\text{Pl}^2}\,\phi_0\>,  
\end{align}
$\phi_d$ as a function of $\phi_0$ is given by 
\begin{align}\label{eq:phid}
   \phi_d = \phi_0 + \frac{\mathfrak{D}_0}{\mu_0^2}\left(1 - \cosh(\mu_0d)\right)\>,
\end{align}
and $\phi_0$ is solution to 
\begin{align}
   \phi_M + \frac{\mathfrak{D}_0}{\mu_0\mu_M}\sinh(\mu_0d) = \phi_0 + \frac{\mathfrak{D}_0}{\mu_0^2}\left(1 - \cosh(\mu_0d)\right)\>.
\end{align}
A prototype solution is plotted in Fig.~\ref{Fig2}. The induced pressure inside the planes is $\displaystyle P = -4.73\times10^{-4}$ pN/cm$^2$ (see section \ref{sec:5}).

These dilaton solutions will be instrumental in analyzing the induced pressure in Casimir experiments.

\section{Dilaton Field of a Sphere}\label{sec:4} 

In this section we consider the dilaton field as caused by a static massive sphere surrounded by vacuum with radius $R$ and a homogenous high density $\rho_S$, which is used for the analysis of the \textit{q}BOUNCE and Lunar Laser Ranging bounds. 

Since neutrons are used in the search for dilatons in \textit{q}BOUNCE experiments and neutron interferometry, it is important to understand their interaction with the dilaton. For a certain parameter regime this interaction between neutron and dilaton becomes strong. In this case, the neutron affects the background dilaton field as generated by the mirrors of the experimental setup in a non-negligible way, which in turn weakens the effect on the neutron, \textit{viz.}~screening of the neutron sets in (see also \cite{Burrage_2021}). 
Since we treat the dilaton as a classical field theory, a consistent description of its coupling to a quantum mechanical system is beyond our reach. Therefore, we employ a semi-classical treatment in which the neutron's probability distribution times its mass acts as the source of the dilaton as defined in Eq.~(\ref{nsc}).

Concerning Lunar Laser Ranging, we approximate the Sun, Earth and Moon as spherical sources of dilatons with homogenous density and employ the solutions derived in this section.

The spherically symmetric field equation is given by
\begin{align}
   \frac{d^2\phi}{dr^2} + \frac{2}{r}\frac{d\phi}{dr} &= V_{\text{eff},\phi}(\phi; \rho)\>,
\end{align}
with the boundary conditions 
\begin{align}
   \phi'(0) &= 0\>, \nonumber\\
   \lim_{r\to\infty}\phi(r) &\to \phi_V\>.
\end{align}
Inside the sphere we have density $\rho_S$. The corresponding minimum of the field value is
\begin{align}
  \phi_S = \frac{m_\text{Pl}}{\lambda}\,W\bigg(\frac{\lambda^2V_0}{A_2\rho_S}\bigg)\>.
\end{align}
We expand
\begin{align}
 V_{\text{eff},\phi}(\phi; \rho_S) &\simeq V_{\text{eff},\phi\phi}(\phi_S; \rho_S)\left(\phi - \phi_S\right) \nonumber\\
  &= \mu_S^2\left(\phi - \phi_S\right)
\end{align}
with
\begin{align}
 \mu_S = \frac{1}{m_\text{Pl}}\sqrt{\lambda^2 V_0\,e^{-\lambda\phi_S/m_\text{Pl}} + A_2\rho_S}\>.
\end{align}
For the field equation we find
\begin{align}\label{eomsi}
   \frac{d^2\phi}{dr^2} + \frac{2}{r}\frac{d\phi}{dr} \simeq \mu_S^2\left(\phi - \phi_S\right).
\end{align}
Introducing the field $\varphi$ 
\begin{align}
   \phi - \phi_S = \frac{\varphi}{r}\>,
\end{align}
Eq.~(\ref{eomsi}) takes on the simpler form
\begin{align}
   \frac{d^2\varphi}{dr^2} \simeq \mu_S^2\,\varphi\>.
\end{align}
From the general solution for $\varphi$ we obtain the particular solution for $\phi$, which is convergent for $r\to0$ and satisfies the boundary condition $\phi'(0) = 0$: 
\begin{align}
   \phi(r) = \phi_S + \mathfrak{C}_I\frac{\displaystyle\sinh(\mu_Sr)}{r}\>,
\end{align}
where $\mathfrak{C}_I$ is some constant to be determined later.

In the vacuum outside the sphere we approximate
\begin{align}
 V_{\text{eff},\phi}(\phi; \rho_V) &\simeq V_{\text{eff},\phi\phi}(\phi_V; \rho_V)\left(\phi - \phi_V\right)  \nonumber\\
  &= \mu_V^2\left(\phi - \phi_V\right),
\end{align}
where $\mu_V$ is given by Eq.~(\ref{muv}).

For the field equation we find
\begin{align}\label{eomsi2}
   \frac{d^2\phi}{dr^2} + \frac{2}{r}\frac{d\phi}{dr} \simeq \mu_V^2\left(\phi - \phi_V\right).
\end{align}
With the field $\varphi$
\begin{align}
   \phi - \phi_V = \frac{\varphi}{r}\>,
\end{align}
the field equation again simplifies to
\begin{align}
   \frac{d^2\varphi}{dr^2} \simeq \mu_V^2\,\varphi\>.
\end{align}
The general solution of which provides the $\phi$ solution convergent for $r\to\infty$ and satisfying the boundary condition $\displaystyle\lim_{r\to\infty}\phi(r) \to \phi_V$ 
\begin{align}
   \phi(r) = \phi_V + \mathfrak{C}_O\frac{e^{-\mu_V (r - R)}}{r}\>,
\end{align}
where $\mathfrak{C}_O$ is some constant to be determined next.

The dilaton field has to satisfy the following boundary conditions at the surface of the sphere 
\begin{align}
   \phi(R_-) &= \phi(R_+)\>, \nonumber\\
   \phi'(R_-) &= \phi'(R_+)\>,
\end{align}
which provides explicit expressions for $\mathfrak{C}_I$ and $\mathfrak{C}_O$. 
After some elementary algebraical manipulations we obtain the relations
\begin{align}
   \mathfrak{C}_I &= \frac{\phi_V - \phi_S}{\cosh(\mu_SR)}\frac{1 + \mu_VR}{\mu_S + \mu_V\tanh(\mu_SR)}\>, \nonumber\\
   \mathfrak{C}_O &= -R\,\frac{1 - \frac{1}{\mu_SR}\tanh(\mu_SR)}{1 + \frac{\mu_V}{\mu_S}\tanh(\mu_SR)}\left(\phi_V - \phi_S\right)\>.
\end{align}
In order to quantify the amount of "screening" of a sphere, we introduce a formfactor, which we interpret as a "screening charge" $\mathfrak Q$. 
The interaction of the sphere with its surrounding is given by its outer field. Clearly, with decreasing radius $R$ the sphere becomes increasingly "unscreened" since the dilaton field inside the sphere decreases towards its minimum throughout the volume of the sphere. On the other hand, for large sphere radii the dilaton field reaches 
its minimum already before the center of the sphere, it is no longer "sourced" by the whole volume of the sphere, which becomes "screened". 

\begin{figure}[h]
\centering
\includegraphics[height=0.5\linewidth]{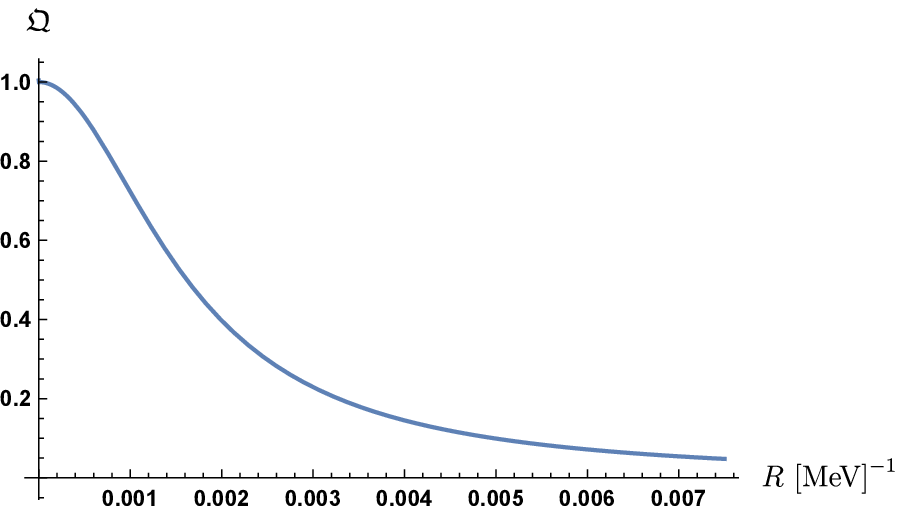} \\
\vspace{5mm}
\includegraphics[height=0.5\linewidth]{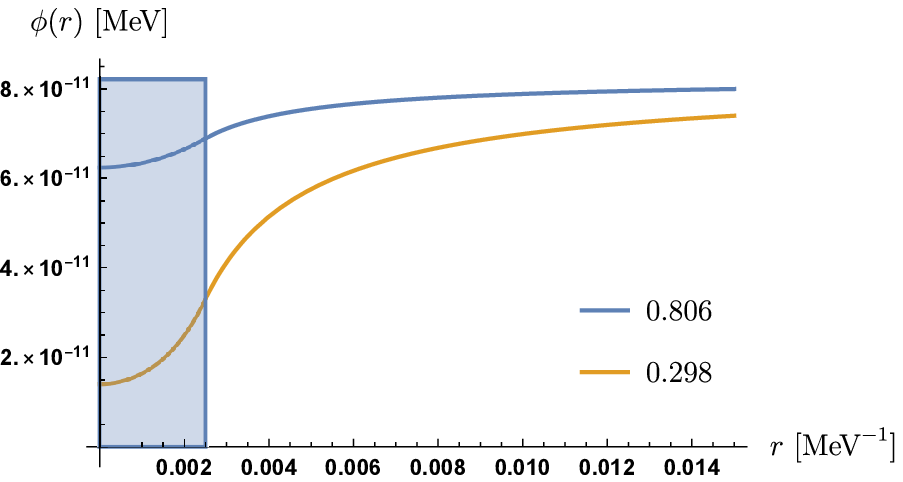} 
\caption{\textit{Top:} The "screening charge" $\mathfrak Q$ is plotted as a function of the radius $R$ of a sphere. The parameters taken are $\rho_N = 1.44\times10^{10}$ MeV$^{4}$, the density of a neutron, $A_2 = 10^{40}$, $\lambda = 10^{15}$, $V_0$ according to Eq.~(\ref{V0AL}) and $R$ varying between $0$ and $3\times R_N = 0.0075$ MeV$^{-1}$, where $R_N$ is the radius of a neutron.
\textit{Bottom:} The field profile of a sphere (neutron) is plotted as a function of the radial distance of the center of the sphere with parameters $A_2 = 10^{39}$, $\lambda = 10^{25}$ and $V_0$ according to Eq.~(\ref{V0AL}). The blue line corresponds to the density of a neutron $\rho_N = 1.44\times10^{10}$ MeV$^{4}$ providing a "screening charge" $\mathfrak Q = 0.806$ and the yellow line to $10\times\rho_N$ providing $\mathfrak Q = 0.298$. The blue square is bounded by the radius $R_N$ of the sphere (neutron) and the vacuum field value $\phi = \phi_V$.}
\label{FigScr}
\end{figure}

An unscreened sphere has $R \ll 1/\mu_S$. Hence, we expand $\mathfrak{C}_O$ in powers of $\mu_SR$ to obtain for an "unscreened" sphere
\begin{align}
   \mathfrak{C}_O^u = -\frac{\mu_S^2R^3}{3}\left(\phi_V - \phi_S\right)\>.
\end{align}
We define the "screening charge" $\mathfrak Q$ as the ratio
\begin{align}\label{scrch}
   \mathfrak Q = \frac{\mathfrak{C}_O}{\mathfrak{C}_O^u} = \frac{3}{\mu_S^2R^2}\frac{1 - \frac{1}{\mu_SR}\tanh(\mu_SR)}{1 + \frac{\mu_V}{\mu_S}\tanh(\mu_SR)}\>.
\end{align}
With this definition, see Fig.~\ref{FigScr},
\begin{align}
  \mathfrak Q \to \begin{dcases}
    0\>, & \quad \text{for "screened" bodies with } \mu_SR \to \infty\>, \nonumber\\
    1\>, & \quad \text{for "unscreened" bodies with } \mu_SR \to 0\>.
  \end{dcases}
\end{align}

Finally, we find for the solution
\begin{align}
  \phi(r) = \begin{dcases}
    &\phi_S + \frac{\phi_V - \phi_S}{\cosh(\mu_SR)}\frac{1 + \mu_VR}{\mu_S + \mu_V\tanh(\mu_SR)} \\
    &\hspace{9mm}\times\>\frac{\sinh(\mu_Sr)}{r}\>, \hspace{26.5mm} \text{for } r\leq R\>, \nonumber\\
    &\phi_V -\mathfrak Q\,\frac{\mu_S^2R^3}{3}\left(\phi_V - \phi_S\right)\frac{e^{-\mu_V(r - R)}}{r}\>, \text{for } r\geq R\>.
\end{dcases}
\end{align}
The acceleration experienced by a pointlike test body, which does not disturb the field, in the outer field of the sphere has been derived in \cite{Pitschmann:2020ejb} for a generic scalar $\phi$ in the non-relativistic limit as
\begin{align}\label{eq:scafor}
  \vec f_\phi = -m\vec\nabla\ln A(\phi)\>.
\end{align}
In all models of consideration $A(\phi) = 1 + \delta A(\phi)$ with $\delta A(\phi)\ll1$, such that $\ln A(\phi) \simeq \delta A(\phi)$. Consequently, we find for the force on a particle caused by a scalar $\phi$ to leading order
\begin{align}\label{eq:SFP}
  \vec f_\phi &= -m\vec\nabla A(\phi)\>.
\end{align}
In the case of dilatons Eq.~(\ref{cf}) holds and we finally obtain for the dilaton force on a point particle 
\begin{align}\label{eq:SFPD}
  \vec f_\phi &= -m\,\frac{A_2}{m_\text{Pl}^2}\,\phi\,\vec\nabla\phi\>,
\end{align}
and, respectively, for the acceleration
\begin{align}
  \vec a_\phi &= -\frac{A_2}{m_\text{Pl}^2}\,\phi\,\vec\nabla\phi\>.
\end{align}
Asymptotically, for large $r$ we obtain
\begin{align}\label{DilaccA}
   \vec a_\phi \to -\mathfrak Q\,\frac{A_2}{m_\text{Pl}^2}\frac{\mu_V\mu_S^2R^3}{3}\,\phi_V\left(\phi_V - \phi_S\right)\frac{e^{-\mu_V(r - R)}}{r}\frac{\vec r}{r}\>,
\end{align}
justifying the definition of $\mathfrak Q$ as a "screening charge". 

These dilaton solutions will be instrumental in analyzing the results of \textit{q}BOUNCE experiment and Lunar Laser Ranging in the following sections. 

\section{Dilaton-induced Frequency Shift in \textit{q}BOUNCE}\label{sec:5}

Here, we derive a discrete set of limits for the \textit{q}BOUNCE experiment \cite{Abele:2009dw, Jenke:2011zz, Jenke:2014yel} using the solutions obtained herein. In this experiment, ultra-cold neutrons are dropped in Earth's gravitational potential and reflected by a neutron mirror, which has been reported for the first time in \cite{Nesvizhevsky:2002ef}. The energy eigenstates are discrete and allow to apply the method of resonance spectroscopy. In \cite{Jenke:2011zz} the basic setup is described. In the Rabi version of the experiment, spectroscopy has been realized with energy resolution 3$\times$10$^{-15}$ peV \cite{Cronenberg:2018qxf}.

The experimental setup is such, that ultracold neutrons pass three regions, while being reflected on polished glass mirrors. In \cite{Cronenberg:2018qxf}, the resonance spectroscopy transitions between the energy ground state $E_1 = 1.41$ peV and the excited states $E_3 = 3.32$ peV as well as $E_4 = 4.08$ peV have been demonstrated. First, the neutrons pass region I which acts as a state selector for the ground state $|1\rangle$ having energy $E_1$. A polished mirror at the bottom and a rough absorbing scatterer on top at a height of about 20 $\mu$m serve to select the ground state. Neutrons in higher, unwanted states are scattered out of the system. This region has a length of 15 cm. Subsequently, in region II, a horizontal mirror performs harmonic oscillations with a tunable frequency $\omega$, which drives the system into a coherent superposition of ground and excited states. The length of this region is 20 cm. Finally, region III is identical to the first region and hence acts again as a ground state selector.

The quantum mechanical description of a neutron above a mirror in the gravitational potential is given by the Schr\"odinger equation \cite{Westphal:2006dj}.
After separation into free transversal and bound vertical states
\begin{align}
   \Psi_n^{(0)}(\textbf{x},t) = \frac{e^{\frac{i}{\hbar}(p_\perp\cdot x_\perp - E_\perp t)}}{2\pi\hbar v_\perp}\,\psi_n^{(0)}(z)\,e^{-\frac{i}{\hbar}E_n t}\>,
\end{align}
it reads 
\begin{align}\label{SEQ}
   -\frac{\hbar^2}{2m_N}\frac{\partial^2\psi_n^{(0)}(z)}{\partial z^2} + m_Ngz\,\psi_n(z) = E_n\psi_n^{(0)}(z)\>.
\end{align}
The characteristic length scale 
\begin{align}
  z_0 = \sqrt[3]{\frac{\hbar^2}{2m_N^2g}} = 5.87\,\mu\text{m}\>,
\end{align}
and energy scale $E_0$ = $\sqrt[3]{\hbar^2 m_Ng^2/2}$ are given by the mass $m_N$ of the neutron and the acceleration of the Earth $g$. With the substitution 
\begin{align}
  \sigma = \sqrt[3]{\frac{2m_N^2g}{\hbar^2}}\left(z - \frac{E_n}{m_Ng}\right) \equiv \frac{z - z_n}{z_0}\>,
\end{align}
Eq.~(\ref{SEQ}) becomes Airy's equation
\begin{align}
   \frac{d^2\tilde\psi_n(\sigma)}{d\sigma^2} - \sigma\,\tilde\psi_n(\sigma) = 0\>.
\end{align}

From the effective dilaton potential
\begin{align}
  V_\text{eff}(\phi) = V_0\,e^{-\lambda\phi/m_\text{Pl}} +  \frac{A_2\rho}{2m_\text{Pl}^2}\,\phi^2\>,
\end{align}
we can deduce the semi-classical neutron-dilaton coupling 
\begin{align}\label{nsc}
   V_\text{eff} = \frac{A_2}{2}\frac{m_N}{m_\text{Pl}^2}\,\psi^*\psi\,\phi^2\>.
\end{align}
There are some subtleties involved here of a nature similar as in the case of the symmetron. We refer to \cite{Brax:2017hna} for further information.
The corresponding quantum mechanical perturbation potential is given by 
\begin{align}\label{symmQMP}
   \textrm{V} = \frac{A_2}{2}\frac{m_N}{m_\text{Pl}^2}\,\phi^2\>,
\end{align}
and leads to a resonance frequency shift to first order  (see e.g. \cite{landau1991quantenmechanik}):
\begin{align}
   \delta E_{mn}^{(1)} &\equiv E_m^{(1)} - E_n^{(1)} \nonumber\\
   &= \frac{A_2}{2}\frac{m_N}{m_\text{Pl}^2}\int_{-\infty}^\infty dz\,\Big(\big|\psi_m^{(0)}(z)\big|^2 - \big|\psi_n^{(0)}(z)\big|^2\Big)\,\phi^2(z)\>.
\end{align}
Likewise, the first order correction to the wavefunctions reads (see e.g. \cite{landau1991quantenmechanik})
\begin{align}
   \Psi_n^{(1)}(\textbf{x},t) = \sum_{m\neq n}\frac{\displaystyle\int d^3x'\,\psi_m^{(0)*}(\textbf{x}')\textrm{V}\psi_n^{(0)}(\textbf{x}')}{E_n^{(0)} - E_m^{(0)}}\,\psi_m^{(0)}(\textbf{x})\,e^{-\frac{i}{\hbar}E_n t}\>.
\end{align}
Hence, the correction to the density $\displaystyle\varrho_n^{(0)}(z)=\psi_n^{(0)*}(z)\psi_n^{(0)}(z)$ to first order is given by 
\begin{align}
   \varrho_n^{(1)}(z) &= 2\,\mathfrak{Re}\big(\psi_n^{(0)*}(z)\psi_n^{(1)}(z)\big) \nonumber\\
   &= A_2\,\frac{m_N}{m_\text{Pl}^2}\,\mathfrak{Re}\sum_{m\neq n}\frac{\displaystyle\int_{-\infty}^\infty dz'\,\psi_m^{(0)*}(z')\,\psi_n^{(0)}(z')\,\phi(z')^2}{E_n^{(0)} - E_m^{(0)}} \nonumber\\
   &\quad\times \psi_n^{(0)*}(z)\psi_m^{(0)}(z)\>,
\end{align}
where $\mathfrak{Re}$ denotes the real part.

In the one mirror case 
the unperturbed normalized wavefunction for $z>0$ reads (see e.g. \cite{Pitschmann:2019boa})
\begin{align}
  \psi_n^{(0)}(z) = \frac{\text{Ai}\Big(\displaystyle\frac{z - z_n}{z_0}\Big)}{\sqrt{z_0}\,\text{Ai}'\Big(-\displaystyle\frac{z_n}{z_0}\Big)}
\end{align}
with $z_n = E_n/(mg)$. 
Outside this region the wavefunction vanishes. 
For a single mirror filling the region $z \leq 0$ we can use Eq.~(\ref{FS1M}) and obtain for the resonance frequency shift
\begin{align}\label{RFS1}
   \delta E_{mn}^{(1)} &= \frac{A_2}{2}\frac{m_N}{m_\text{Pl}^2}\frac{1}{z_0}\int_0^\infty dz\,\Big(\phi_V + \left(\phi_0 - \phi_V\right)e^{-\mu_Vz}\Big)^2 \nonumber\\
   &\quad\times\Bigg\{\frac{\displaystyle\text{Ai}\Big(\frac{z - z_m}{z_0}\Big)^2}{\text{Ai}'\Big(-\displaystyle\frac{z_m}{z_0}\Big)^2} - \frac{\displaystyle\text{Ai}\Big(\frac{z - z_n}{z_0}\Big)^2}{\text{Ai}'\Big(-\displaystyle\frac{z_n}{z_0}\Big)^2}\Bigg\}\>.
\end{align}
It is straightforward to find all the corresponding expressions in the two mirror case. These expressions are very elaborate in their full detail and hence we will refrain from reproducing them herein.

In order to account for the "screening" of the neutron itself for the extraction of the experimental limits, the transition energies should be replaced as follows
\begin{align}
   \delta E_{pq} \to \mathfrak Q\,\delta E_{pq}\>,
\end{align}
where $\mathfrak Q$ is given in Eq.~(\ref{scrch}).
The whole parameter space of dilatons can be efficiently constrained using the results obtained herein. 
The corresponding analysis has been carried out in parallel to this work \cite{Pitschmann:2018aa}, where the solutions obtained here for a one mirror setup are used to exclude regions of the dilaton parameter space.

Nevertheless, in Table \ref{table1} we summarize the resonance frequency shift for a range of dilaton parameters near the experimental sensitivity in the case of a single mirror.
\begin{table}[ht]  
\centering
\renewcommand{\arraystretch}{1.4}
\begin{tabular}{|c||c|c|}
  \hline
  $\delta E_{14}^{(1)}$ [eV] 	& 	$A_2$ 	&	$\lambda$ 	\\
 \hline\hline
$3.03121\times 10^{-21}$ 		& 	$10^{35}$ 			& 	$10^{25}$  \\
\hline
$5.71509\times 10^{-16}$ 		& 	$10^{35}$ 			& 	$10^{30}$ 	 \\
\hline
$5.75507\times 10^{-21}$ 						&	$10^{35}$ 			& 	$10^{35}$  \\
\hline
$2.60699\times 10^{-21}$ 		& 	$10^{40}$ 			& 	$10^{25}$   \\
\hline
$6.76315\times 10^{-12}$ 		& 	$10^{40}$ 			& 	$10^{30}$   \\
\hline
$5.59563\times 10^{-16}$ 		& 	$10^{40}$ 			& 	$10^{35}$ \\
\hline
$1.51542\times 10^{-24}$ 		& 	$10^{45}$ 			& 	$10^{25}$ 	\\
\hline
$1.51550\times 10^{-14}$ 		& 	$10^{45}$ 			& 	$10^{30}$      \\
\hline
$3.02715\times 10^{-13}$ 	& 	$10^{45}$ 			& 	$10^{35}$	  \\
\hline
\end{tabular}
\caption{$\delta E_{41}^{(1)}$ for $1$ mirror, viz. Eq.~(\ref{RFS1}), for the value $\rho = 2.51$ g/cm$^3$.}
\label{table1}
\end{table}
{{As can be seen in this table, the deviation can be larger than the experimental uncertainty of order $3\times 10^{-15}$ eV for cases where for instance $A_2=10^{40}$ and $\lambda=10^{30}$ corresponding to $M\simeq 10^{-2}$ GeV and $\Lambda\simeq 10^{-3}$ eV, i.e. the dark energy scale. As a result the forthcoming results in \cite{Pitschmann:2018aa} will test an interesting range of scales for both particle physics and cosmology.}}

\section{Dilaton-induced Pressure in CASIMIR experiments}\label{sec:6}

Here, we consider limits that can be obtained by the Casimir And Non-Newtonian force EXperiment (\textsc{Cannex}) \cite{Sedmik:2018kqt} (see also \cite{Klimchitskaya:2019fsm, Klimchitskaya:2019nzu}).
This experiment consists of two parallel plates in a vacuum chamber and has been devised to measure the Casimir force and hypothetical fifth forces.
A dilaton field would induce a pressure between those plates, which can be measured with high precision. 

We approximate the setup in one dimension along the $z$-axis as follows. Between the upper surface of the fixed lower mirror at $z = 0$ and the lower surface of the movable upper mirror at $z = d$ there is vacuum. Then follows the upper mirror with thickness $D$ and above that at $z > d + D$ there is vacuum again. 
In order to obtain the induced pressure for the \textsc{Cannex} setup, we can apply the force induced on a point particle by dilatons in Eq.~(\ref{eq:SFP}). We obtain
\begin{align}
  \vec f_\phi = -\rho_M\int_{-\infty}^\infty dx\int_{-\infty}^\infty dy\int_d^{d + D} dz\,\partial_z\ln A(\phi)\,\vec e_z\>.
\end{align}
Consequently, the pressure in $z$-direction is given by
\begin{align}
  P_z = -\rho_M\int_d^{d + D} dz\,\partial_z\ln A(\phi)\>.
\end{align}
The corresponding integral is a surface term and hence trivially carried out with the final result 
\begin{align}
  P_z = \rho_M\,\Big(\ln A\big(\phi(d)\big) - \ln A\big(\phi(d + D)\big)\Big)\>.
\end{align}
This agrees with (see e.g. \cite{Pitschmann:2020ejb})
\begin{align}
  \nabla^\mu T_{\mu\nu} = \partial_\nu\ln A\,T\>,
\end{align}
where $T=g^{\mu\nu}T_{\mu\nu}$ and which reduces for a static field configuration of $\phi$ to
\begin{align}
  \partial^z T_{zz} = -\partial_z P_z  = \partial_z\ln A\,\rho_M\>.
\end{align}
In all models of consideration $A(\phi) = 1 + \delta A(\phi)$ with $\delta A(\phi)\ll1$, such that $\ln A(\phi) \simeq \delta A(\phi)$. For dilatons Eq.~(\ref{cf}) holds and we finally obtain the dilaton-induced pressure
\begin{align}
  P_z = \rho_M\,\frac{A_2}{2m_\text{Pl}^2}\left(\phi^2(d) - \phi^2(d + D)\right).
\end{align}
For $\phi(d)$ we employ the value $\phi_d$ at the mirror surface of the corresponding two mirror solution given in Eq.~(\ref{eq:phid}), while for $\phi(d + D)$ we can use the value $\phi_0$ at the mirror surface of the one mirror solution given in Eq.~(\ref{eq:phi0}) instead. Finally, we obtain for the pressure
\begin{align}
  P_z = \rho_M\,\frac{A_2}{2m_\text{Pl}^2}\left\{\phi_d^2 - \left(\frac{\mu_V\,\phi_V + \mu_M\,\phi_M}{\mu_V + \mu_M}\right)^2\right\}.
\end{align}
This results can be used to carry out a  numerical analysis \cite{Pitschmann:2018aa} where the exclusion regions of the dilaton parameter space are obtained from the \textsc{Cannex} experiment {{as we expect a sensitivity of $\vert P_z\vert \le 1$ nPa.}}

\section{Dilaton Constraints by Lunar Laser Ranging}\label{sec:7} 

In this section we analyze bounds on dilatons due to Lunar Laser Ranging (LLR), which provides three separate constraints \cite{nordtvedt2001lunar} due to 
\begin{enumerate}
\item the Nordvedt effect, which relates to a difference between the free fall acceleration of the Earth and Moon towards the Sun (equivalence principle violations),
\item deviations from the inverse-square law at distances comparable to the Earth-Moon separation and
\item time-variation of $G$.
\end{enumerate}
Since the dilaton interaction is time-independent, it cannot induce a time-variation of the effective coupling $G$. Hence, only the first two constraints can lead to dilaton bounds and will be analyzed here.

\subsection{Constraints due to the Nordvedt Effect}

For the first LLR test we consider the acceleration 
of the Earth (${\earth}$) and Moon (${\leftmoon}$) towards the Sun (${\astrosun}$). 
For the Earth we find \ref{DilaccA}
\begin{align}
\vec a_{\earth} &= \vec a_G + \vec a_{\phi{\earth}} \nonumber\\
&\simeq\vec a_G  - \mathfrak Q_{\astrosun}\mathfrak Q_{\earth}\,\frac{A_2}{m_\text{Pl}^2}\frac{\mu_V\mu_{\astrosun}^2R_{\astrosun}^3}{3}\,\phi_V\left(\phi_V - \phi_{\astrosun}\right)    \nonumber\\
&\quad\times\frac{e^{-\mu_V(r_{AU} - R_{\astrosun})}}{r_{AU}}\frac{\vec r_{AU}}{r_{AU}}\>,
\end{align}
where $\vec r_{AU}$ is the vector pointing from Sun to Earth, $\vec a_G$ is the acceleration due to the Sun's gravitational field and $\vec a_{\phi{\earth}}$ the acceleration due to the dilaton field of the Sun and given by Eq.~(\ref{DilaccA}), which 
holds for a test particle in an outer field without screening. The screening of the Earth is taken into account by the second charge factor $\mathfrak Q_{\earth}$.
Likewise, the acceleration of the Moon is given by 
\begin{align}
\vec a_{\leftmoon} &= \vec a_G + \vec a_{\phi{\leftmoon}} \nonumber\\
&\simeq\vec a_G  - \mathfrak Q_{\astrosun}\mathfrak Q_{\leftmoon}\frac{A_2}{m_\text{Pl}^2}\frac{\mu_V\mu_{\astrosun}^2R_{\astrosun}^3}{3}\,\phi_V\left(\phi_V - \phi_{\astrosun}\right)   \nonumber\\
&\quad\times\frac{e^{-\mu_V(r_{AU} - R_{\astrosun})}}{r_{AU}}\frac{\vec r_{AU}}{r_{AU}}\>,
\end{align}
where $\vec r_{AU}$ is again the vector pointing from Sun to Earth, which is approximately the vector pointing from Sun to Moon. Then, $\vec a_G$ is the acceleration due to the Sun's gravitational field and taken equal for Earth and Moon, such that any difference in acceleration between Earth and Moon, \textit{viz.}~an equivalence principle violation, is attributed to
the dilaton field.
For the equivalence principle violation quantified by the E\"otvos parameter
\begin{align}
\eta_\text{em} = 2\frac{|\vec a_{\earth} - \vec a_{\leftmoon}|}{|\vec a_{\earth} + \vec a_{\leftmoon}|} \simeq \frac{|\vec a_{\phi{\earth}} - \vec a_{\phi{\leftmoon}}|}{|\vec a_G|}\>,
\end{align}
we find in the case of dilatons
\begin{align}
\eta_\text{em} &\simeq\frac{\mathfrak Q_{\astrosun}|\mathfrak Q_{\earth} - \mathfrak Q_{\leftmoon}|}{|\vec a_G|}\,\frac{A_2}{m_\text{Pl}^2}\frac{\mu_V\mu_{\astrosun}^2R_{\astrosun}^3}{3}\,\phi_V\left(\phi_V - \phi_{\astrosun}\right)    \nonumber\\
&\quad\times\frac{e^{-\mu_V(r_{AU} - R_{\astrosun})}}{r_{AU}}\>.
\end{align}
{ {Typically this is constrained at the $\eta_{\rm em}\le 2\times 10^{-13}$ level by the LLR experiment \cite{Williams:2012nc}}}.

\subsection{Constraints due to Deviations from the Inverse-Square Law}

For the second LLR test we consider the precession of the lunar perigee caused by fifth forces. The corresponding Eq.~(\ref{Eq:pf}) is derived in Appendix~\ref{sec:A} and reads
\begin{align}\label{eq:LLR21}
   \frac{\delta\Omega}{\Omega} \simeq - \frac{R^2}{GM_{\earth}}\left(\delta f(R) + \frac{R}{2}\,\delta f'(R)\right).
\end{align}
It is derived in Appendix \ref{sec:A}. Here, $\delta f(R)$ is the acceleration caused by dilatons at the maximum Earth-Moon separation $R$, and given by \ref{DilaccA}
\begin{align}\label{eq:LLR22}
\delta f(R) &\simeq\mathfrak Q_{\earth}\mathfrak Q_{\leftmoon}\frac{A_2}{m_\text{Pl}^2}\frac{\mu_V\mu_{\earth}^2R_{\earth}^3}{3}\,\phi_V\left(\phi_V - \phi_{\earth}\right)   \nonumber\\
&\quad\times\frac{e^{-\mu_V(R - R_{\earth})}}{R}\>.
\end{align}
Using Eq.~(\ref{eq:LLR22}) in Eq.~(\ref{eq:LLR21}) leads to the central relation
\begin{align}
   \frac{\delta\Omega}{\Omega} &\simeq -\mathfrak Q_{\earth}\mathfrak Q_{\leftmoon}\frac{A_2}{m_\text{Pl}^2}\frac{\mu_V\mu_{\earth}^2R_{\earth}^3}{3}\frac{R}{GM_{\earth}}\,\phi_V\left(\phi_V - \phi_{\earth}\right)   \nonumber\\
&\quad\times\left(1 - \frac{\mu_VR}{2}\right)e^{-\mu_V(R - R_{\earth})}\>,
\end{align}
to be compared with experimental results of order $6\times 10^{-12}$.

The results of the numerical analysis for both tests are reproduced in the accompanying paper \cite{Pitschmann:2018aa}.

\section{Conclusion}\label{sec:8} 

We have derived approximate analytical solutions to the dilaton field theory in the presence of a one or two mirror system as well as for a sphere. The 1-dimensional equations of motion have been integrated in each case. The analytical results obtained herein provide the necessary input for the numerical study carried out in parallel to this work \cite{Pitschmann:2018aa}. The latter provides bounds on dilatons by using results from three "experiments". First, from \textit{q}BOUNCE we obtain results by employing bouncing ultracold neutrons, second, \textsc{Cannex} provides bounds by measuring induced pressures between parallel plates and third, from Lunar Laser Ranging we obtain additional bounds in the astrophysical regime.   

\acknowledgments

We thank Hartmut Abele and Ren\'e Sedmik for fruitful discussions. 
P.B. acknowledges support from the European Union’s Horizon 2020 research and innovation programme under the Marie Skłodowska -Curie grant agreement No 860881-HIDDeN. This work was supported by the Austrian Science Fund (FWF): P 34240-N.

\appendix

\section{Precession of the Lunar Perigee induced by Fifth Forces}\label{sec:A}

In this Appendix we derive the precession of the lunar perigee induced by an arbitrary "fifth force". Historically, the following result is essentially due to \textit{Isaac Newton} (\cite{newton1999principia}, Propositions XLIII - XLV). For a modern derivation see \cite{chandrasekhar1995newton}.
In this Appendix, any "force" is understood as the absolute value of the centripetal force per mass
\begin{align}
   f(r) = -\frac{\vec r}{r}\cdot\frac{d\vec v}{dt}\>.
\end{align}
We recall the Binet equation governing $u = 1/r = g(\varphi)$ (see e.g.~\cite{chandrasekhar1995newton})
\begin{align}
  f(r) &= h^2u^2\frac{d^2u}{d\varphi^2} + h^2u^3 \nonumber\\
  &= h^2g^2(\varphi)\frac{d^2g}{d\varphi^2}(\varphi) + h^2g^3(\varphi)\>,
\end{align}
where $h$ is the absolute value of $\vec h = \vec r \times \vec v$.

First, we address the following question. Given that $u = 1/r = g(\varphi)$ is the inverse polar equation of an orbit described under the action of a centripetal force $f(r)$ with a constant of areas $h$, what is the centripetal force $\tilde f(r)$ under which a revolving orbit $u = 1/r = g(\tilde\varphi/\alpha)$ with $\tilde\varphi = \alpha\varphi$ and a constant of areas $\tilde h = \alpha h$ would be described, where $\alpha$ is some assigned constant?
The polar form of an ellipse with its center at one focus and $\varphi$ defined such that $\varphi = 0$ corresponds to the point nearest to this focus, is given by
\begin{align}\label{ellpf}
   \frac{p}{r} = 1 + e\cos\varphi
\end{align}
with eccentricity $e$ and semi-latus rectum $p$. Hence, we have for an ellipse
\begin{align}\label{fell}
   f(r) = \frac{h^2}{p}\frac{1}{r^2}\>.
\end{align}

We now require the centripetal force $\tilde f(r)$ under which the orbit 
\begin{align}
  r = g(\tilde\varphi/\alpha)
\end{align}
will be described with a constant of areas 
\begin{align}
  \tilde h = \alpha h\>.
\end{align}
We find
\begin{align}
  \tilde f(r) &= \tilde h^2u^2\frac{d^2u}{d\tilde\varphi^2} + \tilde h^2u^3 \nonumber\\
  &= \tilde h^2g^2(\tilde\varphi/\alpha)\frac{d^2g}{d\tilde\varphi^2}(\tilde\varphi/\alpha) + \tilde h^2g^3(\tilde\varphi/\alpha) \nonumber\\
  &= \alpha^2 h^2g^2(\varphi)\frac{1}{\alpha^2}\frac{d^2g}{d\varphi^2}(\varphi) + \alpha^2 h^2g^3(\varphi) \nonumber\\
  &= h^2g^2(\varphi)\frac{d^2g}{d\varphi^2}(\varphi) + \alpha^2 h^2g^3(\varphi) \nonumber\\
  &= f(r) - h^2g^3(\varphi) + \alpha^2 h^2g^3(\varphi) \nonumber\\
  &= f(r) - h^2u^3 + \alpha^2 h^2u^3\>, 
\end{align}
and finally \textit{Newton's theorem of revolving orbits} \cite{whittaker1988treatise}
\begin{align}
   \tilde f(r) - f(r) = \frac{\tilde h^2 - h^2}{r^3}\>.
\end{align}

In order to relate fifth forces to the precession of the Lunar perigee, we compare the elliptic orbit described under an inverse-square law of attraction with a nearly circular orbit caused by  an inverse-square law of attraction together with a small fifth force. The elliptic orbit described by Eq.~(\ref{ellpf}) is not revolving, while an additional fifth force leads to a deformation of the orbit and also lets the orbit revolve. Actually, the orbit of the Moon is to first order an ellipse with manifold corrections at higher orders. These higher orders also lead to corrections of the relation between fifth force and precession of the Lunar perigee, which are sub-leading and hence will be neglected. 

For an elliptic orbit we have Eq.~(\ref{fell})
and find for the force causing the revolving orbit
\begin{align}
   \tilde f(r) &= \frac{h^2}{p}\frac{1}{r^2} + \frac{\tilde h^2 - h^2}{r^3} \nonumber\\
   &= \frac{1}{p}\frac{1}{r^3}\left[rh^2 + p\left(\tilde h^2 - h^2\right)\right].
\end{align}
For a nearly circular orbit, we may write 
\begin{align}
   r = R - \delta r\>,
\end{align}
where $R$ is the maximum distance, which for an ellipse is $R = a(1 + e)$. Hence, we find
\begin{align}\label{Pro45a}
   \tilde f(r) = \frac{1}{p}\frac{1}{r^3}\left[\left(R - \delta r\right)h^2 + p\left(\tilde h^2 - h^2\right)\right].
\end{align}
The new force we describe in the form
\begin{align}\label{LLRNF}
   \tilde f(r) = \frac{1}{p}\frac{C(r)}{r^3}\>,
\end{align}
where $C(r)$ is an arbitrary function. Since we consider nearly circular orbits, we perform a Taylor expansion
\begin{align}
   \tilde f(r) \simeq \frac{1}{p}\frac{C(R) - C'(R)\delta r}{r^3}\>,
\end{align}
where $\displaystyle C'(R) = \frac{dC}{dR}$. Eq.~(\ref{Pro45a}) leads to
\begin{align}
   \left(R - \delta r\right)h^2 + p\left(\tilde h^2 - h^2\right) = C(R) - C'(R)\delta r\>.
\end{align}
Comparing coefficients gives
\begin{align}
   Rh^2 + p\left(\tilde h^2 - h^2\right) &= C(R)\>, \nonumber\\
   h^2 &= C'(R)\>.   
\end{align}
For nearly circular orbits we have to leading order $p = R$ and hence
\begin{align}
   R\tilde h^2 &= C(R)\>, \nonumber\\
   h^2 &= C'(R)\>.   
\end{align}
For $\alpha$ we find
\begin{align}
   \alpha = \frac{\tilde h}{h} = \sqrt{\frac{1}{R}\frac{C(R)}{C'(R)}}\>.
\end{align}
Using Eq.~(\ref{LLRNF}) we can express $\alpha$ in terms of $\tilde f(r)$
\begin{align}
   \alpha = \sqrt{\frac{\tilde f(R)}{3\tilde f(R) + R\tilde f'(R)}}\>.
\end{align}

Next, we consider the case of gravitational attraction due to the Earth and an additional small fifth force $\delta f(r)$ acting in radial direction and caused by the Earth. Hence, we have
\begin{align}
   \tilde f(r) = \frac{GM_{{\earth}}}{r^2} + \delta f(r)\>.
\end{align}
To leading order we obtain
\begin{align}
  \alpha \simeq 1 - \frac{R^2}{GM_{{\earth}}}\,\delta f(R) - \frac{1}{2}\frac{R^3}{GM_{{\earth}}}\,\delta f'(R)\>.
\end{align}
The correction to the precession of the Lunar perigee is obviously given by
\begin{align}
   \frac{\delta\Omega}{\Omega} = \alpha - 1\>,
\end{align}
so we can finally connect the fifth force with the precession of the Lunar perigee
\begin{align}\label{Eq:pf}
   \frac{\delta\Omega}{\Omega} \simeq - \frac{R^2}{GM_{{\earth}}}\left(\delta f(R) + \frac{R}{2}\,\delta f'(R)\right).
\end{align}

\newpage

\bibliographystyle{unsrt}
\bibliography{dilaton}

\begin{thebibliography}{10}

\bibitem{Joyce:2014kja}
Austin Joyce, Bhuvnesh Jain, Justin Khoury, and Mark Trodden.
\newblock {Beyond the Cosmological Standard Model}.
\newblock {\em Phys. Rept.}, 568:1--98, 2015.

\bibitem{Khoury:2003rn}
Justin Khoury and Amanda Weltman.
\newblock {Chameleon cosmology}.
\newblock {\em Phys. Rev.}, D69:044026, 2004.

\bibitem{Khoury:2003aq}
Justin Khoury and Amanda Weltman.
\newblock {Chameleon fields: Awaiting surprises for tests of gravity in space}.
\newblock {\em Phys. Rev. Lett.}, 93:171104, 2004.

\bibitem{Brax:2004qh}
Philippe Brax, Carsten van~de Bruck, Anne-Christine Davis, Justin Khoury, and
  Amanda Weltman.
\newblock {Detecting dark energy in orbit: The cosmological chameleon}.
\newblock {\em Phys. Rev. D}, 70:123518, 2004.

\bibitem{Damour:1994zq}
T.~Damour and Alexander~M. Polyakov.
\newblock {The String dilaton and a least coupling principle}.
\newblock {\em Nucl. Phys.}, B423:532--558, 1994.

\bibitem{Babichev:2009ee}
E.~Babichev, C.~Deffayet, and R.~Ziour.
\newblock {k-Mouflage gravity}.
\newblock {\em Int. J. Mod. Phys. D}, 18:2147--2154, 2009.

\bibitem{Brax:2012jr}
Philippe Brax, Clare Burrage, and Anne-Christine Davis.
\newblock {Screening fifth forces in k-essence and DBI models}.
\newblock {\em JCAP}, 1301:020, 2013.

\bibitem{Brax:2014wla}
Philippe Brax and Patrick Valageas.
\newblock {K-mouflage Cosmology: the Background Evolution}.
\newblock {\em Phys. Rev.}, D90(2):023507, 2014.

\bibitem{Vainshtein:1972sx}
A.~I. Vainshtein.
\newblock {To the problem of nonvanishing gravitation mass}.
\newblock {\em Phys. Lett.}, 39B:393--394, 1972.

\bibitem{Pitschmann:2018aa}
Mario Pitschmann, Gunther Cronenberg, Ren\'e Sedmik, Philippe Brax, Christian
  Kaeding, Hauke Fischer, Peter Geltenbort, Tobias Jenke, and Hartmut Abele.
\newblock {Acoustic Rabi oscillations between gravitational quantum states and
  impact on dilaton dark energy}.
\newblock {\em to be published}.

\bibitem{Abele:2009dw}
H.~Abele, T.~Jenke, H.~Leeb, and J.~Schmiedmayer.
\newblock {Ramsey's Method of Separated Oscillating Fields and its Application
  to Gravitationally Induced Quantum Phaseshifts}.
\newblock {\em Phys. Rev.}, D81:065019, 2010.

\bibitem{Jenke:2011zz}
Tobias Jenke, Peter Geltenbort, Hartmut Lemmel, and Hartmut Abele.
\newblock {Realization of a gravity-resonance-spectroscopy technique}.
\newblock {\em Nature Phys.}, 7:468--472, 2011.

\bibitem{Sedmik:2018kqt}
Ren{\'{e}} Sedmik and Phillippe Brax.
\newblock {Status Report and first Light from Cannex: Casimir Force
  Measurements between flat parallel Plates}.
\newblock {\em J. Phys. Conf. Ser.}, 1138(1):012014, 2018.

\bibitem{nordtvedt2001lunar}
Kenneth Nordtvedt.
\newblock Lunar laser ranging --- a comprehensive probe of the post-newtonian
  long range interaction.
\newblock In Claus L{\"a}mmerzahl, C.~W.~Francis Everitt, and Friedrich~W.
  Hehl, editors, {\em Gyros, Clocks, Interferometers...: Testing Relativistic
  Graviy in Space}, pages 317--329, Berlin, Heidelberg, 2001. Springer Berlin
  Heidelberg.

\bibitem{Brax:2010gi}
Philippe Brax, Carsten van~de Bruck, Anne-Christine Davis, and Douglas Shaw.
\newblock {The Dilaton and Modified Gravity}.
\newblock {\em Phys. Rev. D}, 82:063519, 2010.

\bibitem{Brax:2011ja}
Philippe Brax, Carsten van~de Bruck, Anne-Christine Davis, Baojiu Li, and
  Douglas~J. Shaw.
\newblock {Nonlinear Structure Formation with the Environmentally Dependent
  Dilaton}.
\newblock {\em Phys. Rev. D}, 83:104026, 2011.

\bibitem{PhysRevD.97.104044}
Lucila Kraiselburd, Susana~J. Landau, Marcelo Salgado, Daniel Sudarsky, and
  H\'ector Vucetich.
\newblock Equivalence principle in chameleon models.
\newblock {\em Phys. Rev. D}, 97:104044, May 2018.

\bibitem{PhysRevD.84.103521}
Kurt Hinterbichler, Justin Khoury, Aaron Levy, and Andrew Matas.
\newblock Symmetron cosmology.
\newblock {\em Phys. Rev. D}, 84:103521, Nov 2011.

\bibitem{Gasperini:2001pc}
M.~Gasperini, F.~Piazza, and G.~Veneziano.
\newblock {Quintessence as a runaway dilaton}.
\newblock {\em Phys. Rev. D}, 65:023508, 2002.

\bibitem{Damour:2002nv}
T.~Damour, F.~Piazza, and G.~Veneziano.
\newblock {Violations of the equivalence principle in a dilaton runaway
  scenario}.
\newblock {\em Phys. Rev. D}, 66:046007, 2002.

\bibitem{Damour:2002mi}
Thibault Damour, Federico Piazza, and Gabriele Veneziano.
\newblock {Runaway dilaton and equivalence principle violations}.
\newblock {\em Phys. Rev. Lett.}, 89:081601, 2002.

\bibitem{PhysRevD.82.063519}
Philippe Brax, Carsten van~de Bruck, Anne-Christine Davis, and Douglas Shaw.
\newblock Dilaton and modified gravity.
\newblock {\em Phys. Rev. D}, 82:063519, Sep 2010.

\bibitem{Ivanov:2016rfs}
A.~N. Ivanov, G.~Cronenberg, R.~H{\"o}llwieser, M.~Pitschmann, Jenke T.,
  M.~Wellenzohn, and H.~Abele.
\newblock {Exact solution for chameleon field, self-coupled through the
  Ratra-Peebles potential with $n=1$ and confined between two parallel plates}.
\newblock {\em Phys. Rev.}, D94(8):085005, 2016.

\bibitem{Brax:2017hna}
Philippe Brax and Mario Pitschmann.
\newblock {Exact solutions to nonlinear symmetron theory: One- and two-mirror
  systems}.
\newblock {\em Phys. Rev. D}, 97(6):064015, 2018.

\bibitem{Pitschmann:2020ejb}
Mario Pitschmann.
\newblock Exact solutions to nonlinear symmetron theory: {{One}}- and
  two-mirror systems. {{II}}.
\newblock {\em Physical Review D}, 103(8):084013, April 2021.

\bibitem{Brax:2021wcv}
Philippe Brax, Santiago Casas, Harry Desmond, and Benjamin Elder.
\newblock {Testing Screened Modified Gravity}.
\newblock {\em Universe}, 8(1):11, 2021.

\bibitem{Brax:2018iyo}
Philippe Brax, Clare Burrage, and Anne-Christine Davis.
\newblock Laboratory constraints.
\newblock {\em International Journal of Modern Physics D}, 27(15):1848009,
  2018.

\bibitem{Burrage_2021}
Clare Burrage, Benjamin Elder, Peter Millington, Daniela Saadeh, and Ben
  Thrussell.
\newblock Fifth-force screening around extremely compact sources.
\newblock {\em Journal of Cosmology and Astroparticle Physics}, 2021(08):052,
  aug 2021.

\bibitem{Jenke:2014yel}
T.~Jenke et~al.
\newblock {Gravity Resonance Spectroscopy Constrains Dark Energy and Dark
  Matter Scenarios}.
\newblock {\em Phys. Rev. Lett.}, 112:151105, 2014.

\bibitem{Nesvizhevsky:2002ef}
V.~V. Nesvizhevsky et~al.
\newblock {Quantum states of neutrons in the Earth's gravitational field}.
\newblock {\em Nature}, 415:297--299, 2002.

\bibitem{Cronenberg:2018qxf}
Gunther Cronenberg, Philippe Brax, Hanno Filter, Peter Geltenbort, Tobias
  Jenke, Guillaume Pignol, Mario Pitschmann, Martin Thalhammer, and Hartmut
  Abele.
\newblock {Acoustic Rabi oscillations between gravitational quantum states and
  impact on symmetron dark energy}.
\newblock {\em Nature Phys.}, 14(10):1022--1026, 2018.

\bibitem{Westphal:2006dj}
Alexander Westphal, H.~Abele, S.~Baessler, V.~V. Nesvizhevsky, A.~K. Petukhov,
  K.~V. Protasov, and A.~{\relax Yu}. Voronin.
\newblock {A Quantum mechanical description of the experiment on the
  observation of gravitationally bound states}.
\newblock {\em Eur. Phys. J.}, C51:367--375, 2007.

\bibitem{landau1991quantenmechanik}
LD~Landau and EM~Lifshitz.
\newblock {\em Quantenmechanik, Lehrbuch der theoretischen Physik III}.
\newblock Akademie-Verlag Berlin, 1991.

\bibitem{Pitschmann:2019boa}
Mario Pitschmann and Hartmut Abele.
\newblock {Schr\"odinger Equation for a Non-Relativistic Particle in a
  Gravitational Field confined by Two Vibrating Mirrors}.
\newblock 12 2019.

\bibitem{Klimchitskaya:2019fsm}
Galina~L. Klimchitskaya, Vladimir~M. Mostepanenko, Ren{\'{e}} I.~P. Sedmik, and
  Hartmut Abele.
\newblock {Prospects for Searching Thermal Effects, Non-Newtonian Gravity and
  Axion-Like Particles: Cannex Test of the Quantum Vacuum}.
\newblock {\em Symmetry}, 11(3):407, 2019.

\bibitem{Klimchitskaya:2019nzu}
G.~L. Klimchitskaya, V.~M. Mostepanenko, and R.~I.~P. Sedmik.
\newblock {Casimir pressure between metallic plates out of thermal equilibrium:
  Proposed test for the relaxation properties of free electrons}.
\newblock {\em Phys. Rev.}, A100(2):022511, 2019.

\bibitem{Williams:2012nc}
James~G. Williams, Slava~G. Turyshev, and Dale Boggs.
\newblock {Lunar Laser Ranging Tests of the Equivalence Principle}.
\newblock {\em Class. Quant. Grav.}, 29:184004, 2012.

\bibitem{newton1999principia}
Isaac Newton.
\newblock {\em The Principia: mathematical principles of natural philosophy}.
\newblock Univ of California Press, 1999.

\bibitem{chandrasekhar1995newton}
S.~Chandrasekhar.
\newblock {\em Newton's Principia for the Common Reader}.
\newblock Clarendon Press, 1995.

\bibitem{whittaker1988treatise}
Edmund~Taylor Whittaker.
\newblock {\em A treatise on the analytical dynamics of particles and rigid
  bodies}.
\newblock Cambridge University Press, 1988.

\end{thebibliography}

\end{document}